\documentclass[a4paper,11pt]{article}
\usepackage{pos}

\usepackage[noarrows]{ezedits}
\defineEdit{GL}{\color[rgb]{0.5,0.2,0.8}}{\color[rgb]{0.5,0.2,0.8}}
\defineEdit{DGM}{\color[rgb]{0.0,0.60,0.60}}{\color[rgb]{0.0,0.60,0.60}}
\defineEdit{SP}{\color[rgb]{1,0.,0}}{\color[rgb]{1.0,0,0}}

\title{Notes on optimizing a multi-sensor gradient axion-like particle dark matter search}

\emailAdd{gaviland@uni-mainz.de}
\emailAdd{grzegorz.lukasiewicz@doctoral.uj.edu.pl}

\author*[a,b,c]{Daniel Gavilan-Martin}

\affiliation[a]{Johannes Gutenberg-Universit{\"a}t Mainz, 55128 Mainz, Germany}
\affiliation[b]{Helmholtz Institute Mainz, 55099 Mainz, Germany}
\affiliation[c]{GSI Helmholtzzentrum für Schwerionenforschung GmbH, 64291 Darmstadt, Germany}

\author[d,e]{Grzegorz Łukasiewicz}
\affiliation[d]{Marian Smoluchowski Institute of Physics, Jagiellonian University in Krak\'ow, Łojasiewicza 11, 30-348, Krak\'ow, Poland}
\affiliation[e]{Doctoral School of Exact and Natural Sciences, Jagiellonian University in Kraków, Łojasiewicza 11, 30-348, Krak\'ow, Poland}
\author[f]{Derek F. Jackson Kimball}
\affiliation[f]{Department of Physics, California State University – East Bay, Hayward, CA 94542, USA}

\author[d,g]{Szymon Pustelny}
\affiliation[g]{Department of Physics, Harvard University, Cambridge, MA 02138, USA}

\author[a,b,c,h]{Dmitry Budker}

\affiliation[h]{Department of Physics, University of California, Berkeley, CA 94720-7300, United States of America}
 
\author[a,b,c]{Arne Wickenbrock}

\FullConference{2nd Training School and General Meeting of the COST Action COSMIC WISPers  (CA21106) (COSMICWISPers2024)\\
 10-14 June 2024 and 3-6 September 2024\\
Ljubljana (Slovenia) and Istanbul (Turkey)\\}
\abstract{
Axion-like particles (ALPs) arise from well-motivated extensions to the Standard Model and could account for the dark matter.
We discuss the scaling of the sensitivity of a galactic ALP dark matter search with the number of sensors, especially in the ultra-light mass regime, where the measurement time is shorter than the coherence time of the ALP field.
We compare multiple schemes for daily modulated ALP gradient signals, and show that increasing the number of sensors from 1 to 2 improves the signal-to-noise ratio (SNR) by a factor of 2-3. For more than two sensors, the SNR increases as the square root of the number of sensors.
Then, we show that splitting the data into subsets and then averaging its Discrete Fourier Transforms (DFTs) is equivalent to the DFT of the whole dataset in terms of SNR.
}
\begin{document}

\maketitle

\section{Introduction}
The nature of dark matter is one of the most pressing mysteries of modern physics. Light bosons with masses below 1\,eV/$c^2$, where $c$ is the speed of light in vacuum, such as axions or, more generally, Axion-Like Particles (ALPs) are well-motivated candidates that could account for the whole or a fraction of the dark matter. Their unknown mass can range plausibly down to $10^{-22}$\,eV/$c^2$ \cite{jackson_kimball_search_2023}.
In this manuscript, we discuss technical aspects of multi-sensor ALPs searches including stochastic effects. In particular, we are interested on the ultra-light mass range ($ \leq 10^{-13}$\,eV/$c^2$), when the coherence time is longer than the measurement time and the stochastic effects are very significant. 

There are several postulated interactions that can be used to search for ALPs, including interactions with photons, gluons, spins, gravitational waves, etc. \cite{chadha-day_axion_2022,graham_experimental_2015}. Here, we consider the spin interactions with electrons ($e$), protons ($p$), and neutrons ($n$), whose Hamiltonians take the form of
\begin{equation}
    \mathcal{H}_i = g_{aii} \boldsymbol{\nabla}a \cdot \boldsymbol{\sigma}_i\,,
\end{equation}
with $i={e\,,p,\,n}$ and $\boldsymbol{\sigma}$ the spin of the corresponding particle. 

There are many ways ALPs can manifest in a spin-based sensor \cite{afach_what_nodate}. For example, if ALPs are the galactic dark matter, their large occupation number will make them appear in a detector as a stochastic signal \cite{centers_stochastic_2021, lisanti_stochastic_2021}. In such a case, to accurately characterize their signal in a detector, we are interested in the stochastic properties of the ALP field gradient rather than the ALP field itself. This distinction leads to qualitative differences in the phase and signal amplitude fluctuations, which are crucial from the point of view of dark matter searches, in particular those using more than one sensor.

This manuscript is structured in two sections. First, we discuss the increase in sensitivity provided by multi-sensor ALP searches via a derivative coupling. Then, we show that splitting data into subsets and then averaging their Discrete Fourier Transforms (DFTs) is equivalent to performing the DFT of the whole dataset in terms of the Signal-to-Noise Ratio (SNR). The former approach simplifies the elimination of noisy periods of data recording while accomplishing the same sensitivity, which is especially relevant for measurements that extend over several days or months.

\section{The ALP field gradient and sensitivity scaling with number of sensors}
Multi-sensor searches have been proposed and realized; see, for example, Refs.\,\cite{Gavilan-Martin:2024nlo, Safdi_2021_Interferometer,afach_what_nodate,Jiang_2023_search, roberts_2017_search}. Here, we expand upon some ideas already presented in Ref.\,\cite{Gavilan-Martin:2024nlo},
where we discussed a search based on two atomic comagnetometers located around 1000 km apart in Mainz, Germany and Krak\'ow, Poland.
In this context, we focus on the ultra-light regime, where the ALP coherence time $\tau_a$ is longer than the total measurement time $T$. A key aspect of spin interactions is that the stochasticity for a gradient search manifests differently from the stochasticity of the field itself.  As shown in detail in Refs.\,\cite{Gavilan-Martin:2024nlo,lisanti_stochastic_2021}, the ALP field gradient in the ultra-light regime is characterized by Rayleigh-distributed, random, and independent amplitudes $\alpha_x,\, \alpha_y,\,\alpha_z$ in the three orthogonal spatial directions, as well as three distinct random phases $\phi_x,\, \phi_y,\,\phi_z$ following a uniform distribution over $[0,2\pi)$.
It should be noted that sensors that are sensitive to ALP-gradient-spin interactions usually have a sensitive axis. Moreover, when the measurement duration extends over 24 hours, the modulation of the sensitive axis due to Earth's rotation has to be taken into account. Thus, the total ALP signal is a carrier $A$ at the ALP frequency $\omega_a$ and two sidebands $A_\pm$ offset by the  Earth's sidereal frequency, i.e., at $\omega_a\pm\omega_E$ \cite{Gavilan-Martin:2024nlo},
\begin{equation}\label{eq:amplitudes}
    \begin{split}
    &|A| =\frac{g_\text{eff}}{\mu_n} \alpha_z \cos\theta  \, , \\
    &|A_-| =\frac{g_\text{eff}}{\mu_n}\frac{\sin\theta}{2}\sqrt{ \alpha_x^2 + \alpha_y^2 - 2\alpha_x\alpha_y\sin(\phi_x-\phi_y) }\, , \\
    &|A_+|  = \frac{g_\text{eff}}{\mu_n} \frac{\sin\theta}{2}\sqrt{ \alpha_x^2 + \alpha_y^2 + 2\alpha_x\alpha_y\sin(\phi_x-\phi_y) } \, ,
    \end{split}
\end{equation}
where $\theta = \angle(\boldsymbol{\hat{\omega}_E}, \boldsymbol{\hat{m}})$ is the angle between the sensitive axis $\mathbf{\hat{m}}$ and the Earth's rotation axis $\boldsymbol{\hat{\omega}_E}$, $\mu_n$ is the gyromagnetic ratio of the atomic species of the sensor, and $g_\text{eff}$ is the effective coupling of the ALP with the spin of the particle of interest (i.e., electron, neutron, or proton).
In our analysis, we are interested in the total signal amplitude defined as $\sqrt{|A|^2+|A_-|^2+|A_+|^2}$.
By sampling across all orthogonal spatial directions, a multi-sensor approach capable of capturing all $\alpha_i$ amplitudes reduces the likelihood of the total signal amplitude being at the lower end of the distribution during the measurement period. By using Eq.\,(\ref{eq:amplitudes}) and assuming sensors with frequency-independent noise and standard deviation over time $\sigma=1$, that measure an ALP field of amplitude $\alpha_i=1$, we can simulate the SNR for an arbitrary number of sensors with arbitrary orientation combined in the same way as in Ref.\,\cite{Gavilan-Martin:2024nlo}.\footnote{Note that the particular $\alpha$ and $\sigma$ chosen are not relevant, since we are interested in the SNR of the different scenarios with respect to each other and not the absolute value of the SNR.} 
\begin{figure}[htb]
    \centering

        \includegraphics[width=0.5\textwidth]{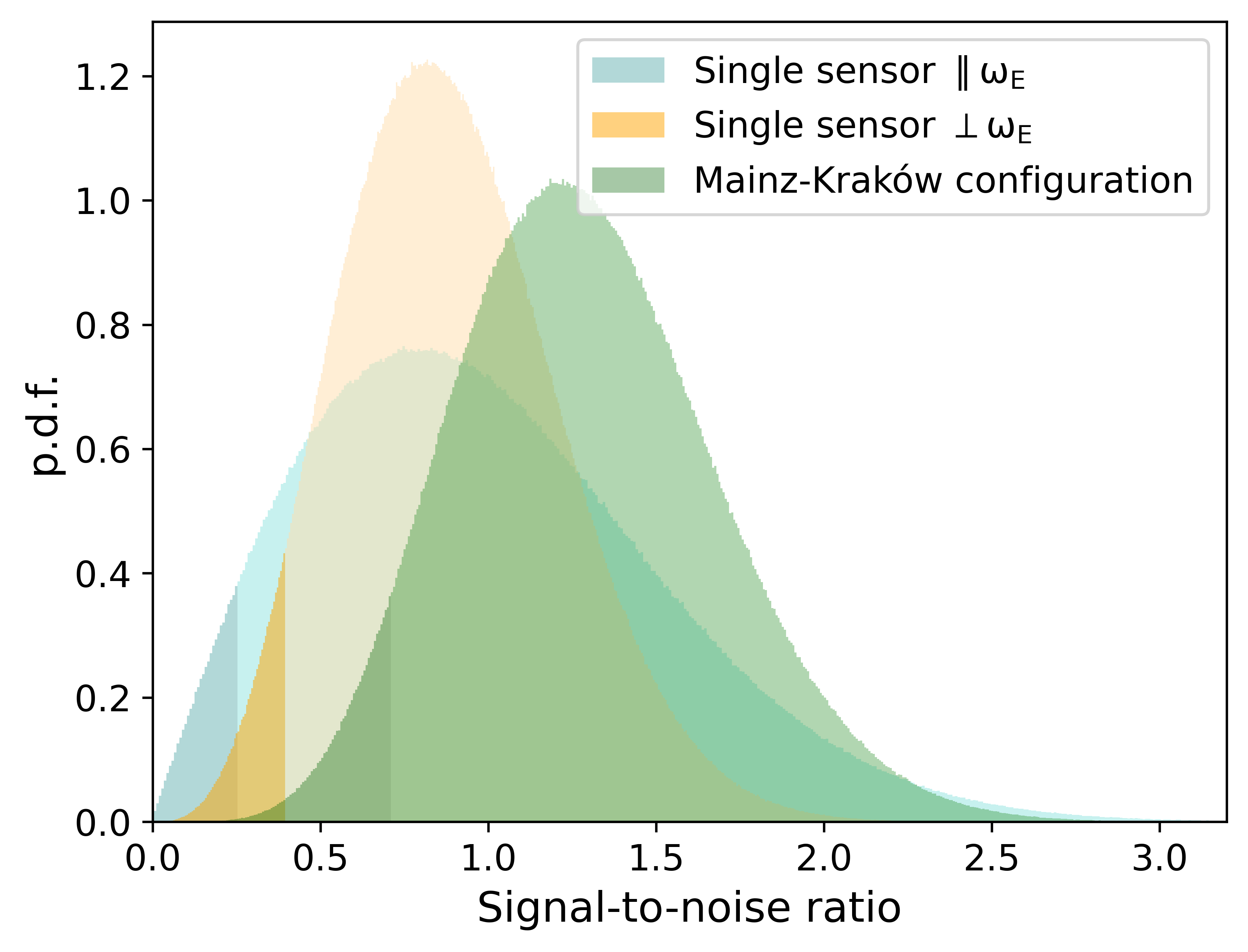}
        
        \caption{Probability distribution functions of the SNR of the ALP gradient signal recorded in the sensors, assuming ALP amplitudes of $\alpha_i=1$ and noise with standard deviation in each detector of $\sigma=1$. Both single-sensor probability distribution functions, parallel and perpendicular to $\boldsymbol{\hat{\omega}_E}$, average to the same value, but the fact that the sensor perpendicular to the Earth's rotation axis can access $\alpha_x$ and $\alpha_y$ decreases the likelihood of a low-probability, extremely low amplitude during the coherence time of the ALP field. 
        The low end of the distribution determines the excluded region when setting limits, since it requires a 95\% confidence level. The 95\% confidence level is displayed as a darker region in each of the p.d.f. 
        }
    \label{fig:pdfSNR}
\end{figure}

Figure \ref{fig:pdfSNR} shows the SNR probability distribution function (p.d.f.) for a single sensor with different orientations with respect to the Earth's rotation axis. It also displays the combined SNR of two sensors in the Mainz-Krak\'ow setup, as described in Ref.\,\cite{Gavilan-Martin:2024nlo}, where the Mainz sensor is oriented at $\theta=90$° and Krak\'ow at $\theta=50$°. The
Mainz-Krak\'ow setup decreases the probability in the lower end of the SNR distribution. For a measurement time shorter than the coherence time, the ALP-field SNR is a single draw from said p.d.f. The SNR is spread more widely over possible amplitudes for a single sensor when oriented parallel to $\boldsymbol{\hat{\omega}_E}$ instead of perpendicular. The sensor perpendicular to $\boldsymbol{\hat{\omega}_E}$ would then cover two orthogonal spatial directions over the course of a day, and hence the chances of being in a low-probability, low-SNR region decrease.

\begin{figure}[htb]
    \centering

        \includegraphics[width=0.5\textwidth]{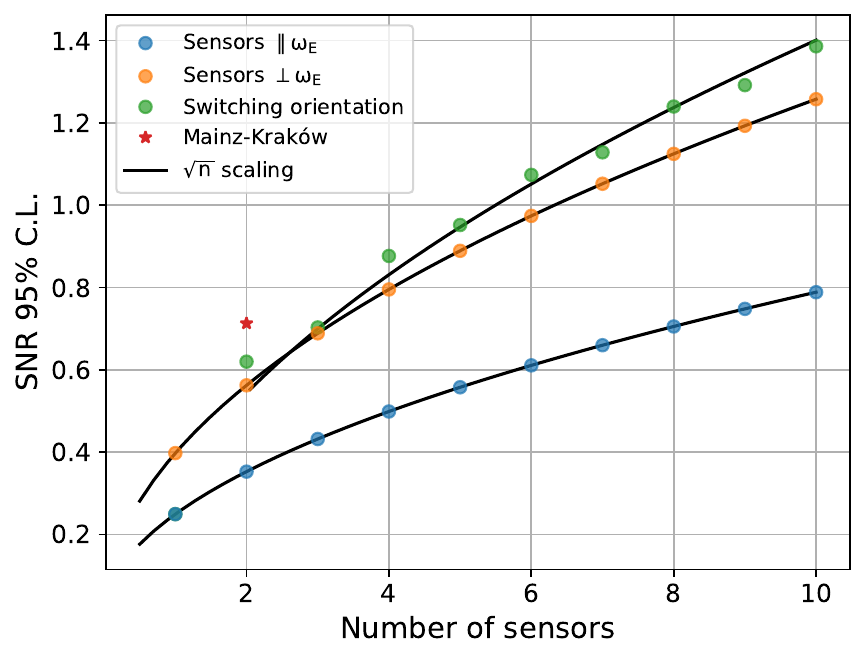}
        
        \caption{Signal-to-noise ratio of the ALP signal at the 95\% confidence level, usually used for setting limits. Adding sensors with a particular orientation, either parallel or perpendicular to $\boldsymbol{\hat{\omega}_E}$, follows a $\sqrt{n}$ scaling. Increasing the number of sensors from one to two with different orientations (like the Mainz-Kraków set up \cite{Gavilan-Martin:2024nlo}) increases the sensitivity by a factor of 2-3.
        Switching the orientation of the newly added sensor (green) between parallel and perpendicular to $\boldsymbol{\hat{\omega}_E}$ is more efficient than using a single orientation, since it allows coverage of the three independent orthogonal directions of the ALP gradient.
        For the alternating sensor scenario, adding more sensors follows a $\sqrt{n}$ scaling, with some fluctuations depending on where the newly-added detector points with respect to $\boldsymbol{\hat{\omega}_E}$. % Greg: I think that single sensor at 45 deg would be very interesting to see here. 
          }
    \label{fig:SensorsvsSNR}
\end{figure}

Figure \ref{fig:SensorsvsSNR} displays the increase in SNR at the 95\% confidence level of the ALP signal as a function of the number of sensors for different orientation scenarios. This value is typically used to set constraints when experimental searches find no evidence of an ALP DM signal.
One might expect a scaling of $\sqrt{n}$ when combining $n$ sensors in an ALP DM search \cite{derevianko_detecting_2018}. However, due to the sampling of the additional spatial independent directions of the ALP gradient, it was demonstrated using simulations, that, in our scheme, the increase reaches a factor of 1.8 (2.8) when going from a single sensor parallel (perpendicular) to Earth's rotation axis $\boldsymbol{\hat{\omega}_E}$ \cite{Gavilan-Martin:2024nlo}. There are some fluctuations in the scaling (as can be seen in the switching orientation scenario in green) depending on the direction of the sensitive axis of the newly added sensors. A sensor perpendicular to $\boldsymbol{\hat{\omega}_E}$ would be sensitive to the two sidebands, in contrast to only the carrier, which is the case for sensors parallel to $\boldsymbol{\hat{\omega}_E}$. Here, as opposed to a single sensor SNR, all spatial directions are already sampled.
In the parallel case, the signal amplitude is divided into two frequency bins corresponding to $A_-$ and $A_+$. When the total signal amplitude is combined, noise is also added, which affects the SNR.

\section{Signal-to-noise ratio scaling with measurement time}
The rule of thumb for the SNR scaling with measurement time $T$ is $\sqrt{T}$. In this section, we will revisit this statement and examine the underlying assumptions. Additionally, we will explore whether the scaling behavior depends on the chosen method of data analysis. Specifically, we will address the differences between performing DFT on the entire signal and averaging multiple DFTs of shorter signal segments. This latter approach is particularly useful for long-duration measurements that may be unintentionally interrupted, making it impractical to directly apply the FFT algorithm to the entire dataset.

Let us consider a discrete time series $x[n] = x_s[n] + \eta[n]$ of length $N$ consisting of a deterministic oscillation $x_s[n] = A\cos(2\pi k_s n/N)$ of amplitude $A$ and frequency index $k_s$, in presence of a noise term following a normal distribution, \hbox{$\eta[n] \sim \mathcal{N}(\mu=0, \sigma)$}. Such a data set can be analyzed in the frequency domain via DFT. We will compare SNR obtained in two analysis schemes: 1)~performing a DFT of the whole data set and 2) averaging DFTs of consecutive data segments in order to manage measurement interruptions. 

\subsection{DFT definition}
Let us first define DFT as 
\begin{equation}
\label{eq:DFT}
    X[k] = \mathrm{DFT}(x)[k] 
    = \frac{1}{N}\sum_{n=0}^{N-1} x[n] e^{-2\pi i nk/N}\,,
\end{equation}
where $N$ is the total number of samples in the date set. Here, we use a normalization factor that is subject to convention. We choose the $1/N$ normalization (so-called \textit{forward}) to comply with the scaling of the power spectrum, where the discrete signal $x_s[n] = A \cos(2\pi k_s n/N)$ has the power of $S[k_s] = 2X_s[k_s]X^*_s[k_s] = A^2/2$, which is independent of the length $N$ of the signal array. In the chosen DFT convention, the amplitude of a harmonic signal at the frequency index $k_s$ is

\begin{equation}
\begin{split}
    X_s[k_s] & = \frac{1}{N}\sum_{n=0}^{N-1} A \cos(2\pi k_s n/N) e^{-2\pi i n\frac{k_s}{N}} 
    = \frac{A}{2N}\sum_{n=0}^{N-1} \left( e^{2\pi i k_s n/N} + e^{-2\pi i k_s n/N} \right) e^{-2\pi i n\frac{k_s}{N}} \\
    & = \frac{A}{2N}\sum_{n=0}^{N-1}(e^0 + e^{-4\pi i k_sn/N}) 
    =  \frac{A}{2N}(N+0) = A/2\,, 
\end{split}
\end{equation}
where $e^{-4\pi i k_sn/N}$ terms sum to zero, because they describe evenly spaced points on a circle in the complex plane. 
Indeed, the power is given by $S[k_s] = 2X_s[k_s]X^*_s[k_s] = A^2/2$, where $X_s^*$ denotes complex conjugate of $X_s$. 

\subsection{DFT -- average or not?}
We have just shown that the absolute value of the complex Fourier amplitude of signal $x_s$ is $A/2$. In order to estimate the SNR, we derive the noise distribution in the frequency domain. The discrete Fourier transform of noise-only time series $\eta$ for index $k$ is given by
\begin{equation}
\label{eq:noise-dft}
    \tilde{\eta}[k] = \frac{1}{N}\sum_{n=0}^{N-1}\eta[n] e^{-2\pi i nk/N}\,,
\end{equation}
where we sum $N$ independent normally distributed random variables $\eta[n]$ with weights given by the complex number $e^{-2\pi i nk/N}$. To evaluate the distribution of $\tilde{\eta}$, let us first consider its real part. Each term in the sum above satisfies $\mathfrak{Re}\left(\eta[n] e^{-2\pi i nk/N}\right) = \eta[n]\cos(-2\pi nk/N)$. Computing the variance of a normally distributed random variable $\eta[n]$ scaled by a constant term $\cos(-2\pi nk/N)$ for each $n$, we find that the variance is given by 

\begin{equation}
    \textrm{Var}\left[\mathfrak{Re}\left(\eta[n] e^{-2\pi i nk/N}\right)\right] = \sigma^2 \cos^2(-2\pi nk/N)\,. 
\end{equation}
Similarly, we obtain the variance of noise-only DFT elements [see Eq.\,(\ref{eq:noise-dft})] by evaluating the variance for a sum of $N$ independent random variables
\begin{equation}
\label{eq:var}
    \textrm{Var}\left[\mathfrak{Re}\left(\tilde{\eta}[k]\right)\right] = \frac{1}{N^2} \sum_{n=0}^{N-1} \sigma^2 \cos^2(-2\pi nk/N) = 
    \frac{\sigma^2}{N^2} \sum_{n=0}^{N-1} \cos^2(-2\pi nk/N) = \frac{\sigma^2}{N^2} \frac{N}{2} = \frac{1}{2N}\sigma^2,
\end{equation}
where the last step is true for $N>2$.

%Since $n$ goes from $0$ to $N-1$, the weights cover the full period of the cosine function with the root-mean-square (RMS) of $1/\sqrt{2}$ (RMS is the correct choice, because the variance of IID is additive and proportional to the square of the underlying random variable centered around 0). Because all values of $\xi[n]$ are IID random numbers centered around 0, it is convenient to substitute all weights with the RMS value. From here it is straightforward to derive the distribution of the white noise time series as the average of $N$ random variables $\xi/\sqrt{2} \sim \mathcal{N}(\mu=0, \sigma/\sqrt{2})$:

The sum of normally distributed random variables is also normally distributed, leading to the conclusion that both real and imaginary parts of each DFT element $\tilde{\eta}[k]$ follow 
\begin{equation}
\begin{split}
    \mathfrak{Re}\left(\tilde{\eta}[k]\right), \mathfrak{Im}\left(\tilde{\eta}[k]\right) \sim \mathcal{N}\left(\mu=0, \sigma/\sqrt{2N}\right) \,,
    \end{split}
\end{equation} 
where the distribution for the imaginary part is derived based on a symmetry argument. 

The SNR can be defined as 
\begin{equation}
\label{eq:SNR}
    \mathrm{SNR} = \frac{A/2}{\sqrt{\textrm{Var}[\tilde{\eta}]}}\,,
\end{equation}
where $\textrm{Var}[\tilde{\eta}] \equiv \textrm{Var}[\mathfrak{Re}(\tilde{\eta}[k])] = \textrm{Var}[\mathfrak{Im}(\tilde{\eta}[k])]$ is independent of the index $k$ as long as the noise $\eta$ is frequency independent. The final expression for the SNR in case 1) (when performing a single DFT on the whole data set) is 
\begin{equation}
    \mathrm{SNR} = \frac{A/2}{\sigma / \sqrt{2N}} = \frac{A_{\textrm{rms}}}{\sigma}\sqrt{N} \,,
\end{equation}
where $A_{\textrm{rms}} = A/\sqrt{2}$. The total measurement time relates to the number of measured points through $T=N/f_s$. We have just proved the known fact that for a sinusoidal oscillation with a constant amplitude in the presence of Gaussian noise the SNR scales as $\sqrt{T}$. 

What would change if we divided the time series into $N/M$ segments, each of length $M$ (where $N$ is an integer multiple of $M$), performed the DFT on each segment separately, and then averaged the resulting DFTs?
 
When using the \textit{forward} scaling of the DFT [see Eq.\,(\ref{eq:DFT})], the amplitude of the oscillating signal is independent of the measurement time (or equivalently, the length of the signal array). In particular, the DFT of the $i$-th $M$-long segment of signal $x_s$ is $(X_s)_i^M = A/2$ and the average over all segments is also $\overline{X_s^M} = A/2$.\footnote{Here we assume that the signal frequency $f_s = k_s/N$ matches also the frequency grid of the $M$-elements long DFT by satisfying $f_s = k_M/M$ for some integer number $k_M$.} Following Eq.\,(\ref{eq:var}), the variance of the DFT values for each analyzed segment is $\textrm{Var}  \left[\mathfrak{Re}\left(\tilde{\eta}^M\right)\right] =\sigma^2/2M$. The average of all $N/M$ segments can be written as \hbox{$\overline{\tilde{\eta}^M} = \frac{1}{N/M}\sum_{i = 1}^{N/M} \tilde{\eta}^M_i $}, leading to the conclusion that the variance of the averaged noise term 
\begin{equation}
    \textrm{Var}\left[\overline{\tilde{\eta}^M}\right] = \frac{1}{(N/M)^2}\sum_{i=1}^{N/M}\frac{\sigma^2}{2M} = \frac{\sigma^2}{2N}
\end{equation}
is invariant under the signal segmentation (independent~of~$M$). 

Although data segmentation does not change the SNR, it is worth noting that the main difference between the two approaches lies in the frequency resolution of the resulting periodograms. In the case of the DFT of the total time series, its frequency resolution is $\Delta f_N = 1/T_N$, where $T_N = N/f_s$ is the measurement time and $f_s$ is the sampling rate. In the second method consisting of averaging DFTs of shorter time series segments of length $M$ the frequency resolution is worse by a factor of $\Delta f_M / \Delta f_N = N/M > 1 $.\footnote{Due to windowing effects, a better frequency resolution is desirable when searching for a signal oscillating at an unknown frequency that might not necessarily match the frequency grid, as in $f_s = k_M /M$, where $k_M$ is an integer.} While segmenting the signal preserves the SNR scaling, it comes at the expense of lower frequency resolution compared to analyzing the full signal.

%It is also important to distinguish data segmentation from signal being coherent over the whole time series. In order to preserve the signal amplitude of $A/2$ the signal needs to be averaged in phase for all available segments. If this is not the case and the signal is only coherent over time $\tau_a < T$, the expected average value of the signal amplitude is attennuated by factor $\sqrt{tau_c/T}$, leading to the same scaling as the noise attenuation with $T$. Naively, after measuring for longer than the coherence time, there is no more gain in SNR. However, a detailed analysis shows that incoherent signal cannot be modeled as a single frequency oscillation and by taking its power spectral distribution into account one finds the correct scaling for incoherent averaging to be $(\tau_a T)^{1/4}$ [see for example the Appendix of \cite{budker_proposal_2014}]. 

It is important to distinguish between the meaning of data segmentation and the coherence of the signal over the entire time series. To preserve the signal amplitude of $A/2$, the signal must be averaged in phase over all available segments. If this is not the case, and the signal is only coherent over a time interval $\tau_a < T$, the expected average value of the signal amplitude will be attenuated by a factor of $\sqrt{\tau_a/T}$ (independent of the chosen analysis method, i.e., with or without segmentation), leading to a scaling similar to that of noise attenuation with $T$. Naively, after measuring for a time longer than the coherence time, there would be no further gain in the SNR. %However, a detailed analysis reveals that an incoherent signal cannot be modeled as a single-frequency oscillation. 
Taking into account the power spectral distribution of an incoherent signal, the correct scaling is found to be $(\tau_a T)^{1/4}$ \hbox{(see Appendix 5 in Ref. \cite{budker_proposal_2014})}.

\section{Conclusion}
We have discussed the scaling of the sensitivity of an ALP DM search with the number of sensors and for different sensitive-axis orientations. Due to the stochastic properties of the ALP gradient, it is beneficial in terms of detection power to cover all three orthogonal directions of the gradient of the ALP field in the considered model. 

With our results, we have shown that segmenting the time series does not affect the scaling of the SNR, provided the segments can be coherently averaged. At the same time, segmenting the signal can be beneficial for managing interruptions in long-duration measurements, but it comes at the expense of reduced frequency resolution compared to analyzing the entire signal. In this context, choosing the right approach requires balancing the trade-off between resilience to interruptions and the desired frequency resolution.

\section{Acknowledgements}
We acknowledge support by the German Research Foundation (DFG) within the German Excellence Strategy (Project ID 39083149); by COST Action COSMIC WISPers CA21106, supported by COST (European Cooperation in Science and Technology). SP acknowledges the support from the National Science Center, Poland, within the OPUS program (2020/39/B/ST2/01524) and GL acknowledges support from the Excellence
Initiative – Research University of the Jagiellonian University in Krak\'ow.
The work of DFJK was supported by the U.S. National Science Foundation under grant PHYS-2110388.
\bibliographystyle{JHEP}
\bibliography{Bibliography}

\end{document}